\documentclass[aip,graphicx]{revtex4-1}

\usepackage{graphicx}
\usepackage{url}
\usepackage{color}

\draft

\begin{document}
\title{Characterization of the pressure coefficient of manganin and temperature evolution of pressure in piston-cylinder cells}
\author{Li Xiang}
\email[]{ives@iastate.edu}
\author{Elena Gati}
\author{Sergey L. Bud'ko}
\affiliation{Ames Laboratory, Iowa State University, Ames, Iowa 50011, USA}
\affiliation{Department of Physics and Astronomy, Iowa State University, Ames, Iowa 50011, USA}
\author{Raquel A. Ribeiro}
\affiliation{Department of Physics and Astronomy, Iowa State University, Ames, Iowa 50011, USA}
\affiliation{CCNH, Universidade Federal do ABC (UFABC), Santo Andr\'{e}, SP, 09210-580, Brazil}
\author{Arif Ata}
\author{Ulrich Tutsch}
\author{Michael Lang}
\affiliation{Physikalisches Institut, Goethe-Universit\"{a}t Frankfurt, Max-von-Laue-Strasse 1, 60438 Frankfurt, Germany}
\author{Paul C. Canfield}
\affiliation{Ames Laboratory, Iowa State University, Ames, Iowa 50011, USA}
\affiliation{Department of Physics and Astronomy, Iowa State University, Ames, Iowa 50011, USA}
\email[]{canfield@ameslab.gov}

\date{\today}

\begin{abstract}
	We report measurements of the temperature- and pressure-dependent resistance, $R(T,p)$, of a manganin manometer in a $^4$He-gas pressure setup from room temperature down to the solidification temperature of $^4$He ($T_\textrm {solid}\sim$ 50\,K at 0.8\,GPa) for pressures, $p$, between 0\,GPa and $\sim\,$0.8\,GPa. The same manganin wire manometer was also measured in a piston-cylinder cell from 300\,K down to 1.8\,K and for pressures between 0\,GPa to $\sim\,$2\,GPa. From these data, we infer the temperature and pressure dependence of the pressure coefficient of manganin, $\alpha(T,p)$, defined by the equation $R_p = (1+\alpha p) R_0$ where $R_0$ and $R_p$ are the resistance of manganin at ambient pressure and finite pressure, respectively. Our results indicate that upon cooling $\alpha$ first decreases, then goes through a broad minimum at $\sim$\,120\,K and increases again towards lower temperatures. In addition, we find that $\alpha$ is almost pressure-independent for $T\,\gtrsim$\,60\,K up to $p\,\sim\,$2\,GPa, but shows a pronounced $p$ dependence for $T\,\lesssim$\,60\,K. Using this manganin manometer, we demonstrate that $p$ overall decreases with decreasing temperature in the piston-cylinder cell for the full pressure range and that the size of the pressure difference between room temperature and low temperatures ($T\,=\,1.8\,$K), $\Delta p$, decreases with increasing pressure. We also compare the pressure values inferred from the magnanin manometer with the low-temperature pressure, determined from the superconducting transition temperature of elemental lead (Pb). As a result of these data and analysis we propose a practical algorithm to infer the evolution of pressure with temperature in a piston-cylinder cell. 
\end{abstract}
\maketitle 
\section{Introduction}
Pressure, as an external tuning parameter, has been recognized as a powerful tool to modify materials' properties as well as to stabilize new, and sometimes exotic, phases\cite{Schilling1979,Doniach1977,Chu2009,Paglione2010,Steglich2016,Mao2016,Drozdov2015,Gati2020a}. To put materials under pressure, a sample of interest is typically placed into a pressure cell surrounded by a pressure-transmitting medium (gas, liquid or solid powder). When a force is applied to the medium via a piston or an anvil, pressure is generated and transmitted to the sample. Over decades, various pressure cells were developed to cover different pressure ranges and many measurement techniques were adapted to be used in these cells\cite{Bridgman1952,Jamieson1962,Smith1966,JayaramanRSI1967,Fujiwara1980,Eremets1996,Colombier2007,Rueetschi2007,Drozdov2015}. In the area of high-pressure research, it is essential to determine the absolute value of the applied pressure that a material of interest is exposed to. Intuitively, assuming some level of hydrostaticity of the pressure medium, one can calculate the pressure $p$ by $p = \frac{F}{S}$, where $F$ is the applied force and $S$ is the area the force is applied to. However, this method suffers from the ambiguity of determination of the exact experienced force and area (due to friction and dimension changes of the area which the force applied to). In addition, the pressure in many pressure cells, particularly in clamp cells, is subject to temperature-induced changes due to differential thermal expansion of the cell materials and media \cite{Brandt1974,Eiling1981,Thompson1984}. Due to these uncertainties, the absolute value of pressure is instead determined from measurements of a physical quantity of a reference system (manometer) where the pressure dependence of the specific physical quantity is pre-characterized. For example, pressure can be determined from measuring the $p$ dependence of the superconducting transition temperature $T_\textrm c$ of elemental Pb, Sn and In\cite{Smith1967,Smith1969,Clark1978,Wittig1979,Bireckoven1988}, the $p$ dependence of the resistance of manganin\cite{Bridgman1911}, the $p$ dependence of the fluorescence lines of ruby (typically used in pressure cells with access for optical measurements, e.g., in diamond anvil cells)\cite{Forman1972,Barnett1973,Piermarini1975}, the $p$ dependence of the lattice parameters of Au, Cu and Pt (often used in neutron or x-ray diffraction experiments)\cite{Fei2007}. The choice of the manometer in a specific experiment often depends on the specific cell design as well as the available measurement techniques.

Among the different types of pressure cells, piston-cylinder clamp cells are among the most commonly used due to their relative ease of usage, their wide covered pressure range (up to $\sim$\,4 GPa, depending on the specific design and materials) as well as their relatively large sample volumes that allow to perform a variety of measurements\cite{Kadomatsu1979,Fujiwara1980,Kamishima2001,Fujiwara2007,Gati2019,Gati2020}. In these cells, either superconducting manometers (Pb, Sn or In) are frequently used to determine pressure at low temperatures or resistive manganin sensors are often utilized to infer pressure at different temperatures, given its relatively high, relatively temperature-insensitive and relatively pressure-sensitive resistivity. Using these sensors, several studies were performed to determine the pressure change as a function of temperature in piston-cylinder cells with maximum pressure of 2 - 3\,GPa\cite{Itskevich1964,Brandt1974,Eiling1981,Thompson1984,Fujiwara2007,Becker1976}. Overall, these studies suggested a pressure drop up to $\sim$\,0.3\,GPa - 0.4\,GPa from room temperature to low temperatures, with some differences in details of $p(T)$ behavior\cite{Itskevich1964,Brandt1974,Eiling1981,Thompson1984,Fujiwara2007}. Some of these estimates \cite{Fujiwara2007,Brandt1974,Thompson1984} relied on the characterization of the pressure-dependent resistance of the manganin sensor at room temperature to obtain the pressure coefficient $\alpha$, defined via $R_p=(1+\alpha p)R_0$ where $R_0$ and $R_p$ are resistance at ambient pressure and finite pressure $p$, respectively. The room-temperature $\alpha$ was then extended to be used at lower temperatures. In fact, other studies suggested already that $\alpha$ is slightly temperature-dependent and therefore the use of a temperature-independent $\alpha$ would result in an overestimation of the pressure change with temperature \cite{Itskevich1964,Andersson1997,Dmowski1999}. Specifically, Dmowski \textit{et al.} in Ref. \onlinecite{Dmowski1999} carried out a temperature-dependent study of $\alpha$ in the $T$ range from 77\,K up to 350\,K. They reported that $\alpha$ decreases linearly with $T$ from 77\,K up to 110\,K, then shows a very sharp change of slope and increases linearly with $T$ up to high temperatures. Despite the fact that for many modern complex materials and phenomena there is a need to accurately evaluate pressure behavior not only at room temperature or liquid Helium temperatures, but also at intermediate temperatures\cite{Kaluarachchi2017PRBa,Xiang2018,Lamichhane2018}, the temperature dependence of $\alpha$ of manganin has not been widely appreciated and used in investigations of the detailed temperature evolution of pressure in piston-cylinder cells. 

The goal of this study is to perform a more detailed and careful characterization of the temperature and pressure dependence of the coefficient $\alpha(T,p)$ of manganin, as well as to utilize it to determine the evolution of pressure with temperature in a piston-cylinder cell. To this end, we first present an analysis of the manganin wire resistance from measurements performed inside a $^4$He-gas pressure cell, which serve as calibration measurements of the manganin sensor, between 0\,GPa and 0.8\,GPa, from room temperature down to the solidification temperature of $^4$He ($T_\textrm {solid}\,\sim\,$50\,K at 0.8\,GPa). In this set of experiments, we make use of the fact that the specific design of the $^4$He-gas pressure setup allows us to readily measure the pressure at low temperatures via a manganin pressure sensor, which is held at room temperature at all times, as long as the pressure medium $^4$He is either in its gaseous or liquid state (see below for more details). In a second step, the resistance of the same manganin wire manometer was measured in a piston-cylinder cell from 300\,K down to 1.8\,K and for pressures between 0\,GPa and $\,\sim$2\,GPa. By combining the results of these measurements, the pressure coefficient, $\alpha (T,p)$, is obtained. We find that $\alpha$ shows a non-monotonic behavior as a function of temperature with a broad minimum at $\sim$\,120\,K. We also show that whereas for $T\,\gtrsim\,$60\,K $\alpha$ is almost pressure-independent, it has a larger pressure dependence for $T\,\lesssim$\,60\,K. Overall, our results emphasize the need to take the temperature and pressure dependence of $\alpha$ into account when using manganin as a secondary manometer. By using the determined $\alpha(T,p)$, we then address the change of pressure with temperature in a piston-pressure cell. We find (i) that pressure decreases with decreasing temperature for all investigated pressures up to $\sim\,$2\,GPa, and (ii) that the pressure difference between room temperature and base temperature, $\Delta p$, decreases with increasing pressure. For our specific combination of pressure cell, pressure medium and sample space filling factor, $\Delta p$ is estimated to be $\simeq$\,0.47\,GPa ($\simeq\,$0.26\,GPa) for lowest (highest) pressure, for which the pressure at low temperature is $\,\simeq\,$0.21\,GPa ($\,\simeq\,$1.86\,GPa). We also compare the pressure values from the manganin sensor at $T\,\simeq\,7\,$K to those, determined from the superconducting transition temperature of elemental Pb (denoted in the manuscript as Pb-$T_\textrm c$ manometer). As a result of this analysis, we offer in the end a ``practical'' approach for inferring $p$ values for our piston-cylinder cell, pressure medium and sample space filling factor for temperatures below room temperature. We note that in previous studies\cite{Eiling1981} the absolute resistance of Pb was also proposed to be used as a manometer for higher temperatures (referred to as Pb-resistive manometer). As we describe in detail in Appendix B, it turns out that the determination of pressure values from a Pb-resistive sensor is somewhat fraught with problems related to the residual resisitity of Pb and the reproducibility of ambient-pressure resistivity values, and therefore a comparison to those values is not included in the main text.

\section{Experimental Details}
The studied manganin manometer was made from a commercial, AWG 44 manganin wire segment (Driver-Harris Co). It has a diameter of $\sim$\,0.05\,mm and was wound into a free-standing coil with an outer diameter of $\sim$\,1.5\,mm. Prior to taking all data, presented here, the manganin manometer was thermally cycled between 300\,K and 1.8\,K for more than ten times under different pressures up to 2\,GPa. After this thermal cycling process, no significant further aging effect of the manganin wire was observed at room temperature. Specifically, the resistance of manganin at room temperature and ambient pressure was the same within 0.01$\%$ before and after a pressure cycle up to $\sim$ 2\,GPa . The Pb manometer was made in-house from elemental Pb with purity higher than 99.99$\%$. In a first step, a 0.03\,mm thin Pb sheet was formed by rolling a glass vial over the elemental Pb piece. Then a rectangular Pb bar with dimensions around 0.7$\times$0.1$\times$0.03\,mm$^3$ was cut from the Pb sheet for electrical resistance measurements.

Resistance measurements of manganin were carried out in the $^4$He-gas pressure setup under pressure up to $\sim$\,0.8\,GPa upon cooling in a $^4$He VTI cryostat down to 5\,K with a cooling rate of -0.2\,K/min. A standard four-terminal configuration was used. Contacts for manganin were made by soldering 100\,$\mu$m diameter Cu wires using a Sn:Pb-60:40 alloy. The manganin wire was supplied with a constant DC current of 10\,$\mu$A and the resulting voltage was measured using a Keithley 2182A Nanovoltmeter. The current direction was switched once during each measurement to subtract thermoelectric voltage contributions. The pressure cell is manufactured out of CuBe (Unipress, Institute of High-Pressure Physics, Polish Academy of Sciences, Unipress Equipment Division) and is connected via a CuBe capillary (outer/inner diameter: 3\,mm/0.3\,mm) to a Helium-gas compressor (Unipress), which is held at room temperature, during the entire time of the experiment. The gas compressor is not only used for changing the pressure in the system, but also acts as a large gas reservoir to ensure, to a good approximation, that pressure inside the pressure cell is held constant during temperature sweeps. The pressure is measured by a manganin sensor inside the compressor (calibrated by Unipress), which measures the pressure in the entire system (low-temperature pressure cell, capillary and compressor) and is itself not subject to any temperature changes. Throughout the manuscript, we will refer to the pressure value determined from this compressor manometer.

The exact same manganin wire that was measured in the $^4$He-gas pressure system, together with a piece of Pb was mounted into a CuBe/NiCrAl hybrid piston-cylinder cell (abbreviated in the manuscript as PCC) similar to the one described in Ref. \onlinecite{Budko1984}, which has a maximum pressure of $\sim$\,2.5\,GPa. Standard four-terminal resistance measurements were performed in a Quantum Design Physical Property Measurement System (PPMS) on warming with a rate of 0.25\,K/min and with a current excitation of 1\, mA for manganin and 5\,mA for Pb. Contacts for Pb were made by spot-welding 25\,$\mu$m Au wires to the sample. A 4:6 mixture of light mineral oil: n-pentane was used as the pressure medium, which solidifies in the range of 3-4\,GPa at room temperature\cite{Torikachvili2015}. Pressure was changed at room temperature and locked by tightening the top lock-nut. 


\section{Results and Discussion}

\subsection{$^4$He-gas pressure cell measurements}
\label{sec:A}
	
\begin{figure}
	\includegraphics[width=0.7\textwidth]{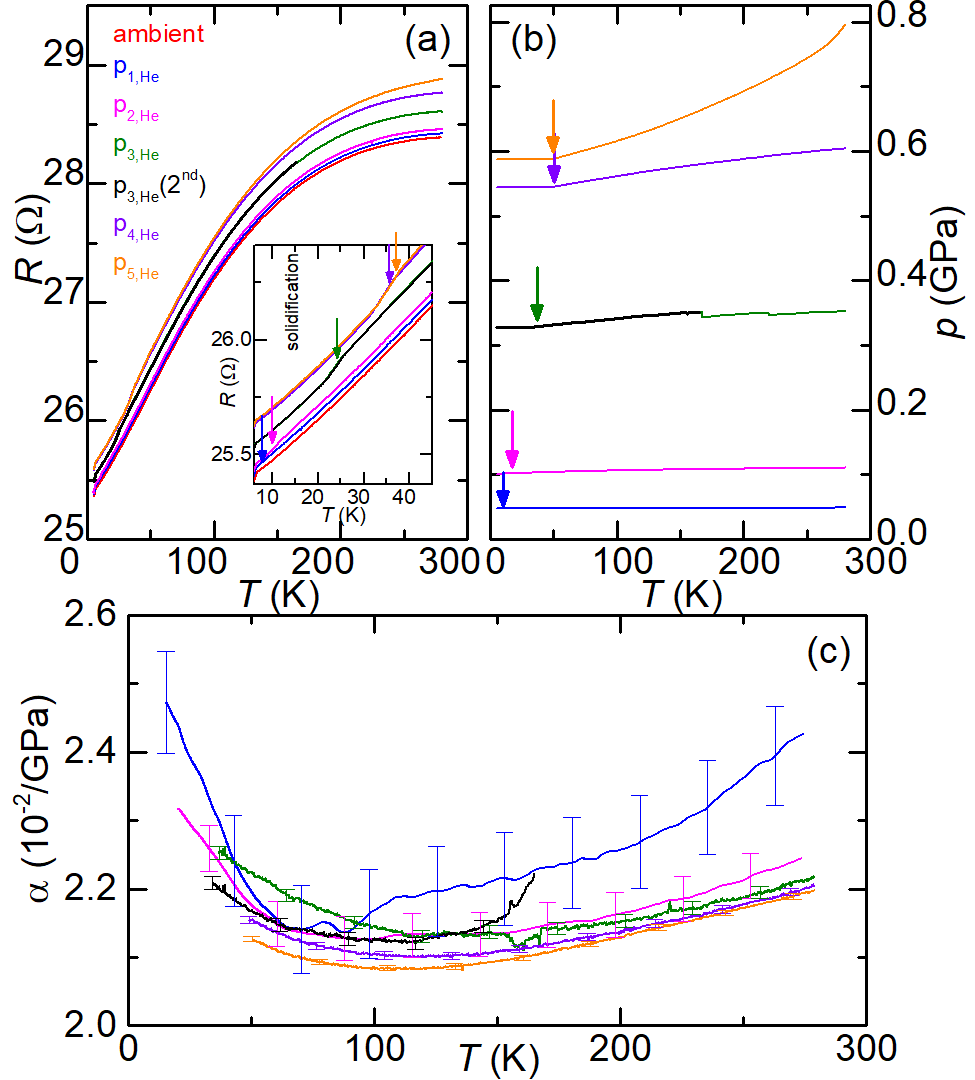}%
	\caption{(a) Temperature-dependent resistance, $R(T)$, of manganin measured in various pressure runs up to $\approx$ 0.8\,GPa in a $^4$He-gas pressure system. The data are labeled according to their run number $\textrm p_\textrm {1,He}$ to $\textrm p_\textrm {5,He}$; Inset: Enlarged view of the low-temperature $R(T)$ data. The kink-like anomalies in $R(T)$ (see arrows) are associated with the solidification of $^4$He; (b) Temperature-dependent pressure, $p(T)$, for the corresponding pressure runs. Pressure values are determined from a manganin manometer, which is located inside the compressor and held at room temperature (see text for details). The step-like change of pressure at $T\,\approx\,160\,$K and 230\,K for the $\textrm p_\textrm {3,He}$ run can be attributed to manual pressure increase via the compressor; (c) Temperature-dependent pressure coefficient, $\alpha(T)$, for various pressure runs. Error bars are a result of an uncertainty of $\pm$0.002\,GPa in the pressure  deternimation of $^4$He-gas pressure system. Data below $^4$He-solidification temperature are discarded due to reasons outlined in the main text.
		\label{fig1}}
\end{figure}

The resistance of the manganin wire, which acts as a secondary manometer, was characterized in a $^4$He-gas pressure cell under pressure up to $\sim$\,0.8\,GPa. Figure\,\ref{fig1}\,(a) presents the temperature-dependent resistance, $R(T)$, of manganin for different pressure runs, denoted as $\textrm p_\textrm {i,He}$, i=1, ...,5. At any temperature, $R$ increases with increasing pressure, and in any pressure run, $R$ decreases with lowering temperature. For all finite pressure runs, kink-like anomalies were observed at low temperatures. The positions of the anomalies (see arrows in Figs.\,\ref{fig1}\,(a) and (b)) are pressure-dependent and can be associated with the solidification of $^4$He\cite{Pinceaux1979}. The temperature dependence of the pressure in the $^4$He-gas experiments (see Fig.\,\ref{fig1}\,(b)), which was recorded by the compressor manometer, shows that the pressure varies only weakly with temperature; this is enabled by the large gas reservoir, provided by the compressor (Note that a leak in the gas-pressure system was responsible for the strong temperature dependence of $\textrm p_\textrm {5,He}$)\cite{Manna2012}. The minor temperature dependence for $\textrm p_\textrm{1,He}$ to $\textrm p_\textrm {4,He}$ can be rationalized when considering that the gas reservoir volume is large but finite ($V\,\sim$\,1000\,cm$^3$ with piston in lowest positions; for comparison cell volume $V\,\sim$\,1\,cm$^3$). Upon cooling, helium atoms are transferred from the reservoir to the pressure cell, leading to an overall minor decrease of the pressure in the entire system with lowering the temperature. As can be intuitively understood from a consideration of ideal gas law, the change of pressure with temperature becomes slightly larger upon decreasing pressure (as seen in Fig.\,\ref{fig1} (b), d$p$/d$T$ becomes larger upon cooling). In addition, the volume of the gas reservoir is reduced by increasing the absolute pressure of the system, since the piston (in the various pressure stages) is moved to different positions. Thus, temperature-induced changes of the pressure are larger for higher pressures than for lower pressures. As can be seen in Fig.\,\ref{fig1}\,(b), these intuitive expectations (d$p$/d$T$ becomes larger upon decreasing $T$ and/or increasing $p$) are met in our measurements of the $p(T)$ landscape. We would like to stress though, that these effects are fully taken into account in our analysis, since we measure the pressure \textit{in situ} at any temperature. Only when the pressure medium becomes solid at very low temperatures\cite{Pinceaux1979}, the compressor and the pressure cell are decoupled since the solid $^4$He in the capillary blocks the pressure transmission from the reservoir to the pressure cell, and thus, the compressor manometer does not measure the low-temperature pressure (see the plateau in $p(T)$ in Fig.\,\ref{fig1}\,(b), particularly clearly for p$_\textrm{4,He}$ and p$_\textrm{5,He}$). We therefore refrain from including data below the solidification in our analysis.

With the data presented in Figs.\,\ref{fig1}\,(a) and (b), the temperature-dependent pressure coefficient can be calculated via 
\begin{equation}
\alpha (T) = \frac{\frac{\Delta R_p}{R_0}}{p} = \frac{(R_p(T)-R_0(T))/R_0(T)}{p}
\label{eq:alpha}
\end{equation}
where $R_0(T)$ and $R_p(T)$ are the resistances measured at ambient pressure and finite pressure $p$, respectively. The resulting $\alpha$ values as a function of temperature for various pressure runs are shown in Fig.\,\ref{fig1}\,(c). Our calculated $\alpha$ value at room temperature is consistent with previous literature reports of $\alpha(300\,\textrm K)\,=\,$(2.35\,$\pm$\,0.15)$\times10^{-2}$/GPa\cite{Wang1967,Zeto1969,Fujioka1978,Andersson1997,Dmowski1999}. For all pressure runs, the overall behavior of $\alpha(T)$ displays a moderate decrease upon cooling in the high-temperature region and then a increase in the low-temperature region with a broad minimum centered around 120\,K. For high temperatures, the $\alpha$ values determined from pressure runs $\textrm p_\textrm {2,He}$ to $\textrm p_\textrm {5,He}$ agree with each other very well, whereas the $\alpha$ values for $\textrm p_\textrm {1,He}$ are clearly larger than the ones from other runs. We speculate that this deviation is related to the fact that the pressure and pressure-induced resistance changes for $\textrm p_\textrm {1,He}$ are so low that systematic errors in the determination of $\alpha$ are larger. 

\begin{figure}
	\includegraphics[width=0.7\textwidth]{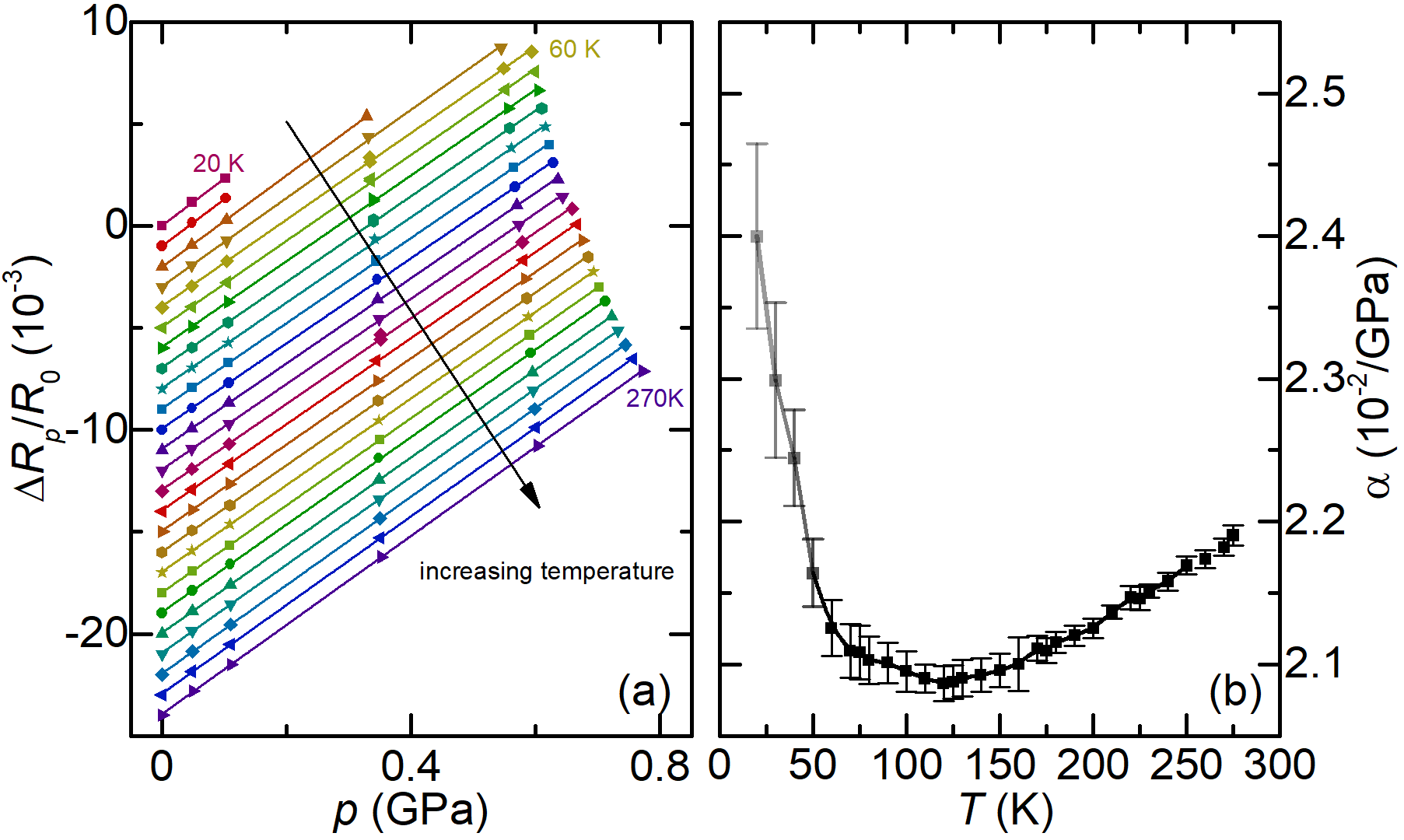}%
	\caption{(a) Normalized change of resistance, defined as $\frac{\Delta R_p}{R_0} = \frac{R_p-R_0}{R_0}$ where $R_0$ and $R_p$ are the resistance at ambient pressure and finite pressure $p$ respectively, as a function of pressure, $p$, for various temperatures from 20\,K to 270\,K (spacing of 10\,K), determined in $^4$He-gas pressure experiments. Lines are linear fits to the data points. Data curves are vertically shifted (spacing of $10^{-3}$) for clarity. For lower temperatures $T\,<\,60\,K$, the high-pressure data points are omitted due to the solidification of the pressure medium (see text); (b) Temperature-dependent pressure coefficient, $\alpha(T)$, obtained by the slope of the linear fit in (a). Error bars correspond to the fitting error of the linear fit. A color gradient for the symbols is used to visualize that the data points result from fitting the data over different pressure ranges, since the solidification of the pressure medium strongly limits the maximum pressure for low temperatures. Black (light grey) symbols indicate that the linear fit was performed up to $\sim$\,0.7\,GPa ($\sim$\,0.2\,GPa). 
		\label{fig2}}
\end{figure}


The temperature dependence of $\alpha$ can be quantified alternatively by analyzing the isothermal pressure dependence of the resistance. Figure\,\ref{fig2}\,(a) presents the normalized change of resistance, $\frac{\Delta R_p}{R_0}$, (defined in Eq.\,\ref{eq:alpha}) as a function of $p$, as determined from our measurements under $^4$He-gas pressure. Up to $\sim$\,0.8\,GPa, $\frac{\Delta R_p}{R_0}$ changes linearly with $p$ for 60 K\,$\leq\,T\,\leq\,$270\,K, i.e. $\alpha$ is constant with $p$ within $5\%$. For $T\,<$\,60\,K, the limited number of data points does not allow us to make a definitive statement on the linearity of $\frac{\Delta R_p}{R_0}$ with $p$ over a wide pressure range. Based on the assumption of linearity, the $\alpha (T)$ data set is obtained by performing a linear fit of the $\frac{\Delta R_p}{R_0}(p)$ data and the result is shown in Fig.\,\ref{fig2}\,(b). The error bars are determined from the error of the linear fits. We relate the larger error bars for $T\,<\,$60\,K to the fact that less data points are available to perform the linear fit. The overall behavior of $\alpha$ as a function of $T$ resembles the data shown in Fig.\,\ref{fig1}\,(c) on a gross level. Note that a non-linear behavior of the $\frac{\Delta R_p}{R_0}(p)$ curve would indicate a $p$ dependence of $\alpha$. In this case, the $\alpha$ value determined from a linear fit of $\frac{\Delta R_p}{R_0}(p)$ data represents an averaged $\alpha$ value over the fitted pressure range, which can be different from the real $\alpha$ value at a specific pressure.

Compared to literature results on the $T$ dependence of the pressure coefficient $\alpha$, our $\alpha(T)$ behavior is overall consistent with that reported in Ref.\,\onlinecite{Dmowski1999} in the sense that a local minimum of $\alpha(T)$ is observed at $T\,\sim\,$120\,K, suggesting that this could be a general behavior of manganin sensor. However, our results suggest a smooth, continuous change of $\alpha$ with temperature, in contrast to the sharp kink anomaly in $\alpha(T)$ at $T\,\approx\,$110\,K as reported in Ref.\,\onlinecite{Dmowski1999}.

\subsection{Piston-cylinder cell measurements} 

\begin{figure}
	\includegraphics[width=0.7\textwidth]{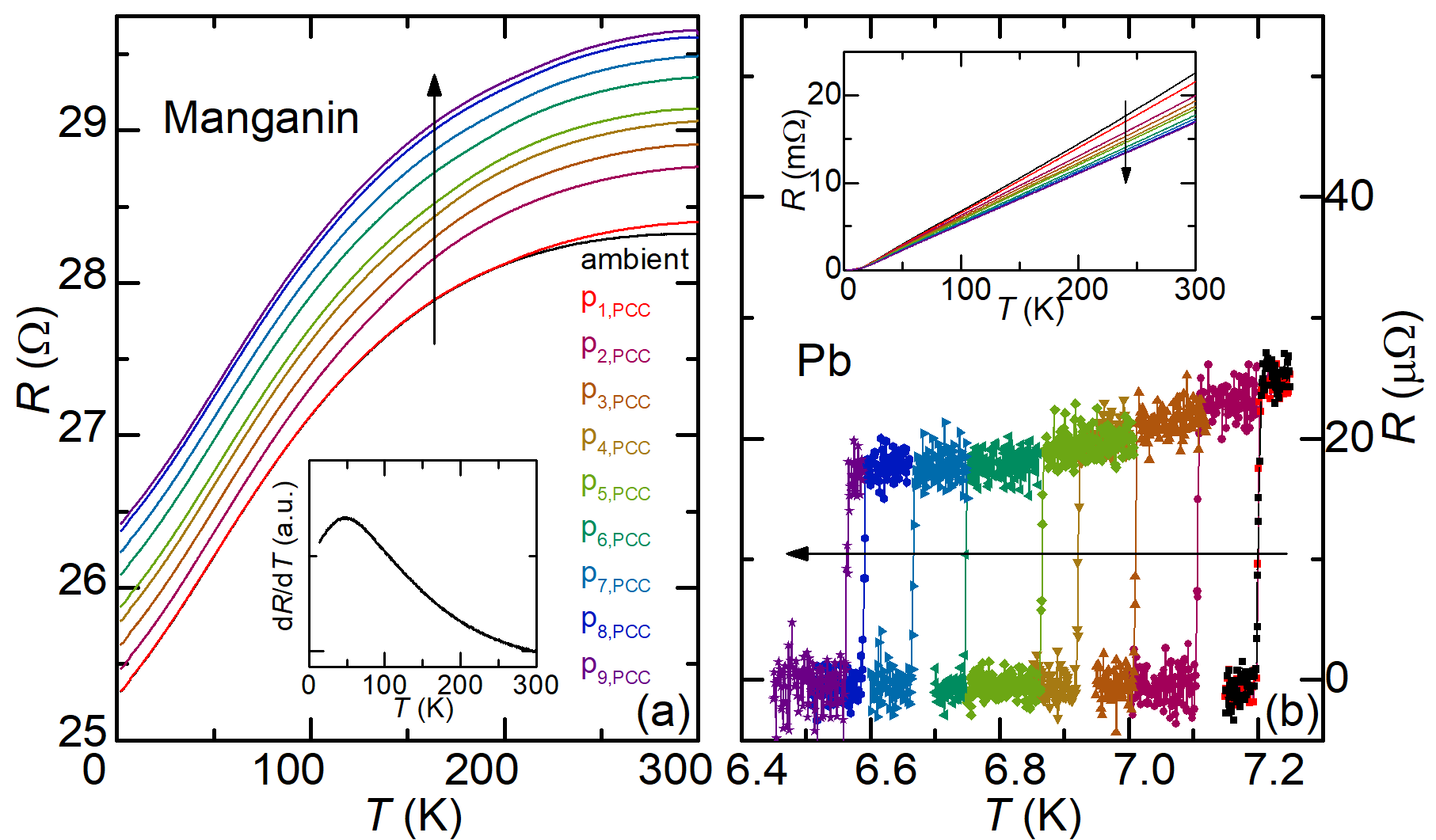}%
	\caption{(a) Temperature-dependent resistance, $R(T)$, of the manganin for different pressure runs up to $\sim$\,2\,GPa measured in a piston-cylinder cell with 4:6 mixture of light mineral oil: n-pentane as a pressure-transmitting medium. Inset: temperature derivative of the manganin resistance, d$R$/d$T$, as a function of temperature at ambient pressure; (b) Temperature-dependent resistance around the superconducting transition of Pb measured in the same experiment. Inset: Temperature-dependent resistance of elemental Pb over the whole temperature range of 1.8\,K - 300\,K. Arrows in the figure indicate the direction of pressure increase.
		\label{fig3}}
\end{figure}

Having obtained a calibration of our manganin sensor from the $^4$He-gas pressure measurements, we proceed and evaluate the temperature dependence of the applied pressure in a piston-cylinder pressure cell. To this end, the characterized manganin sensor, together with Pb-$T_\textrm c$ manometer, is utilized to study the pressure behavior in the PCC. Figure\,\ref{fig3} presents the temperature-dependent resistance of manganin (Fig.\,\ref{fig3}\,(a)) and Pb (inset of Fig.\,\ref{fig3}\,(b)) for various pressure runs up to $\sim$ 2\,GPa. The pressure runs $\textrm p_\textrm{2,PCC}$ - $\textrm p_\textrm{9,PCC}$ were taken after the application of a force, ranging from 1000 lbs to 8000 lbs, by a hydraulic press, whereas for $\textrm p_\textrm{1,PCC}$ the lock-nut was closed hand-tight without the application of external load. The analysis of the $\textrm p_\textrm{1,PCC}$ data will be discussed in Appendix A. As shown in Fig.\,\ref{fig3}, at any fixed temperature $R$ of manganin increases upon increasing pressure. The superconducting transition temperature, $T_\textrm c$, of Pb (Fig.\,\ref{fig3}\,b) is suppressed upon increasing pressure. 


\begin{figure}
	\includegraphics[width=0.7\textwidth]{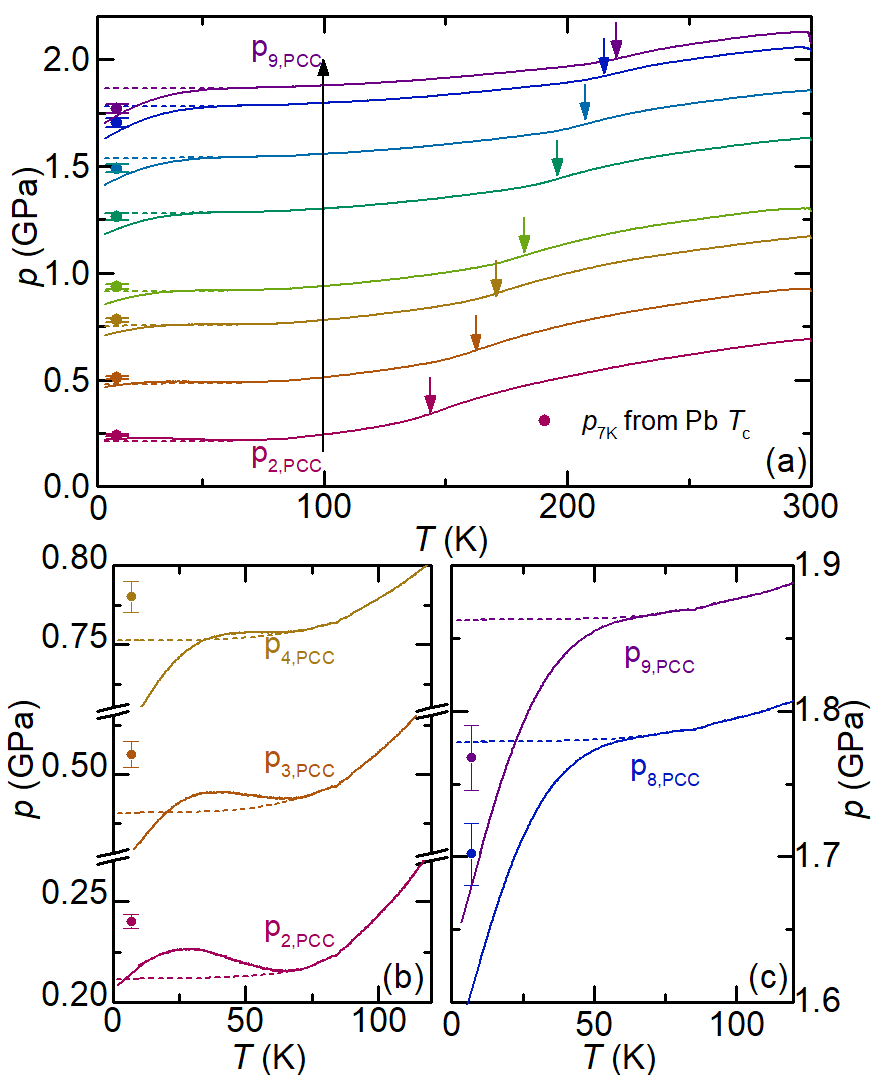}%
	\caption{(a) Temperature-dependent pressure, $p(T)$, for various pressure runs in piston-cylinder cell. Pressure run $\textrm p_\textrm{1,PCC}$ (hand tight) is discussed separately in Appendix A. Solid lines are $p(T)$ curves determined from $\alpha (T)$ obtained from measurements in $^4$He-gas cell (Fig.\,\ref{fig2}\,(b)) and $R(T)$ of manganin measured in PCC (Fig.\,\ref{fig3}\,(a)). Dashed lines correspond to $p(T)$ curves that were extrapolated from high temperatures and represent a physically reasonable $p(T)$ behavior at low temperatures (for details, see text). Circles correspond to pressure values at low temperature, $p_\textrm{7\,K}$, determined from $T_\textrm c$ of Pb. Downward arrows indicate a more rapid pressure decrease in $p(T)$ curves which is associated with the solidification of the pressure medium (see text for details); (b, c) Enlarged view of the low-temperature data of $p(T)$ for lowest pressures  (b) and highest pressures (c).
		\label{fig4}}
\end{figure}

 The $p$ values over the full temperature range from 300\,K down to low temperature are calculated from the manganin resistance using the pressure coefficient $\alpha(T)$ obtained from the $^4$He-gas pressure experiments (see Fig.\,\ref{fig2}\,(b)). The resulting $p(T)$ curves are shown in Fig.\,\ref{fig4} by solid lines. Upon cooling from high temperature, $p(T)$ decreases, until at a certain temperature, which depends on the pressure, a pronounced feature (kink) in $p(T)$ occurs (as shown, e.g., by the arrows at 140\,K for $\textrm p_\textrm{2,PCC}$ or at 220\,K for $\textrm p_\textrm{9,PCC}$ in Fig.\,\ref{fig4}\,(a)). This feature is associated with the solidification of the pressure medium, since its temperature coincides with previous reports on the solidification temperature of the chosen medium\cite{Torikachvili2015}. Upon further decreasing temperature below the solidification, $p(T)$ still continues to decrease, however the slope, d$p$/d$T$, becomes progressively reduced.

Below $T\,\sim$\,60\,K a second set of distinct features appears in $p(T)$, as shown in Fig.\,\ref{fig4}\,(a). In detail, for low pressures ($\textrm p_\textrm{2,PCC}$ to $\textrm p_\textrm{4,PCC}$) $p(T)$ displays a non-monotonic temperature dependence with local minima and maxima below 60\,K (see Fig.\,\ref{fig4}\,(b) for enlarged view), and for higher pressures ($\textrm p_\textrm{5,PCC}$ to $\textrm p_\textrm{9,PCC}$) $p(T)$ shows a rapid decrease below $\,\sim$\,60\,K upon cooling (see Fig.\,\ref{fig4}\,(c) for enlarged view of $\textrm p_\textrm{8,PCC}$ and $\textrm p_\textrm{9,PCC}$). In contrast to the solidification temperature, the temperature of 60\,K does not correspond to any characteristic temperature of the system, since there is, to the best of our knowledge, no drastic change of thermal expansion of any of the cell components \cite{Swenson1997,Ventura2014}. Also, since thermal expansion is typically smaller at lower temperatures and is zero at 0\,K, it is reasonable to assume that the change of pressure with temperature should become smaller for low temperatures and should smoothly change from a finite d$p$/d$T$ for finite temperatures to d$p$/d$T\,=\,$0 at $T\,=\,$0\,K. In the following, we will argue that the features in $p(T)$ below $\sim\,$60\,K in Fig.\,\ref{fig4}\,(a) can be attributed to a non-negligible pressure dependence of $\alpha$ for low temperatures, which for simplicity has been ignored in the analysis so far.

To this end, we construct $p(T)$ curves below 60\,K, which are modified in such a way that they represent a physically more reasonable behavior, and then discuss their implication on the pressure dependence of $\alpha$. For this construction, we used a simple form of polynomial that simultaneously meets the following criteria: (i) the fit describes our experimental $p(T)$ data for 70\,K\,$<\,T\,<\,$90\,K, (ii) the fit reaches d$p$/d$T$\,=\,0 at 0\,K and (iii) d$p$/d$T$ of the fit is always positive. We found that these criteria can be best met by using a polynomial of the order of 4 of the form $p(T) = aT^4+b$, where $a$ and $b$ are fitting parameters. These fits are shown by the dashed lines in Fig.\,\ref{fig4}.

\begin{figure}
	\includegraphics[width=0.7\textwidth]{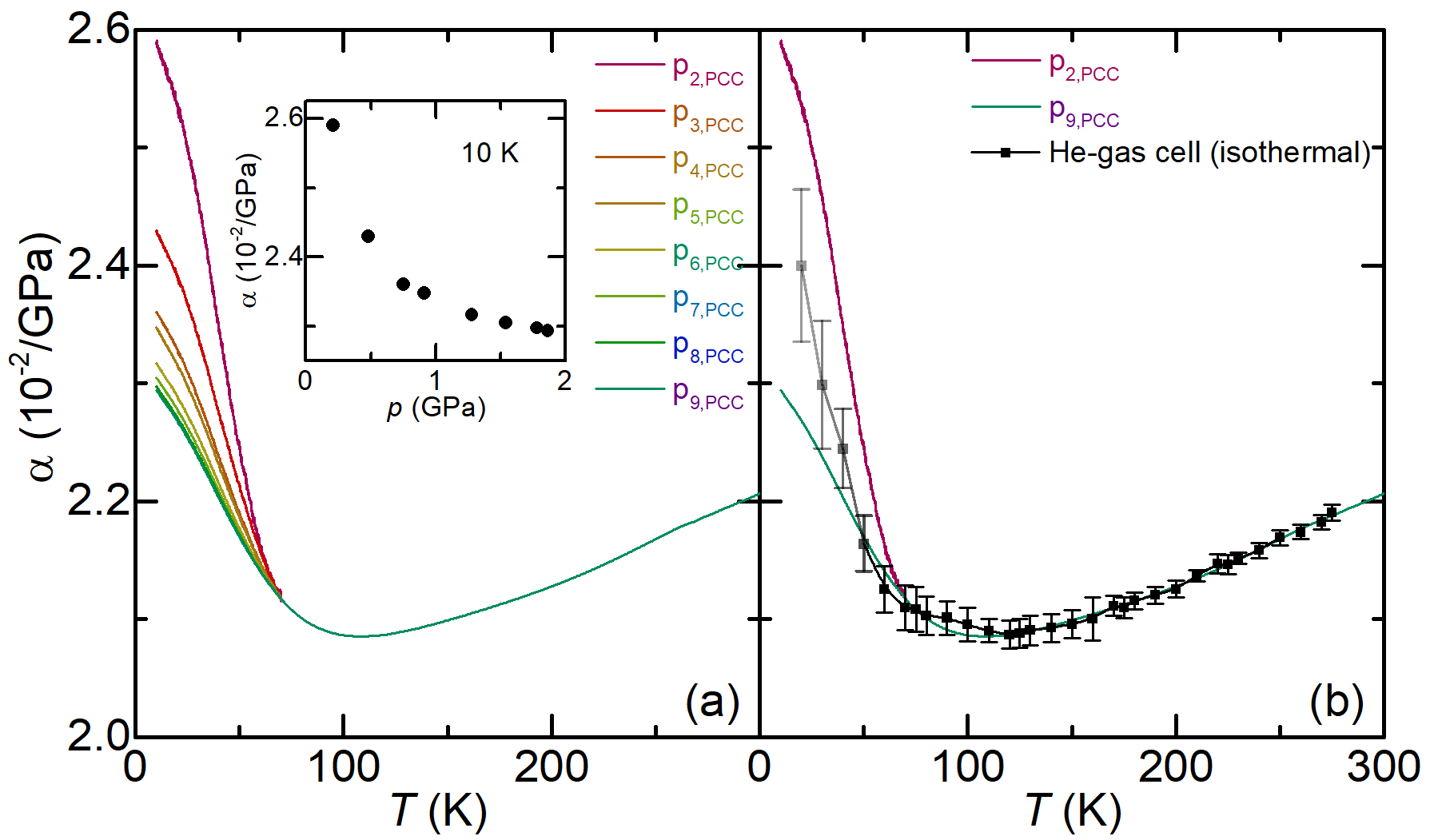}%
	\caption{(a) The temperature-dependent pressure coefficient, $\alpha(T)$, of manganin determined from modified $p(T)$ in the piston-cylinder cell. Inset: The pressure dependence of $\alpha$ at $T$ = 10 K where pressure values at 10 K are obtained from the modified $p(T)$. (b) Comparison of the $\alpha(T)$ determined from the modified $p(T)$ with that determined from $^4$He-gas pressure cell measurements via linear fit of $p$-dependent $\frac{\Delta R_p}{R_0}$ data (same plot as in Fig.\,\ref{fig2}\,(b)). A color gradient for the symbols is used to visualize that the data points result from fitting the $^4$He-gas pressure data over different pressure ranges, as explained in caption of Fig.\,\ref{fig2} and the main text.
		\label{fig5}}
\end{figure}

We can now crosscheck what the implications of our extrapolations of the $p(T)$ behavior for $T\,<\,$60\,K are for the behavior of $\alpha(T,p)$. As shown in Fig.\,\ref{fig5}, the corresponding modified $\alpha(T)$ curves for various pressure runs in the piston-cylinder cell are plotted as lines and symbols. The modified $\alpha(T)$ curves at low temperatures agree with that determined from $^4$He-gas pressure cell measurements on a qualitative level, since for all pressure runs in the piston-cylinder cell $\alpha$ increases rapidly upon cooling below $60$ K, and quantitatively, since the absolute values are within a similar range (see Fig.\,\ref{fig5}\,(b)). As a result of modifying the $p(T)$ behavior at low temperatures, $\alpha$ shows a clear pressure dependence for low temperatures. For any temperature below $\sim\,$60 K, $\alpha$ determined from the modified $p(T)$ in the piston-cylinder cell is suppressed upon increasing pressure. Specifically, $\alpha(\textrm{10 K})$ is suppressed from 2.59$\times 10^{-2}$/GPa to 2.29$\times 10^{-2}$/GPa when the low-temperature pressure is increased from 0.21\,GPa to 1.86\,GPa (see Fig.\,\ref{fig5} (a) inset). Overall, this corresponds to a change of $\alpha$ up to 12$\%$ with pressure at low temperatures, which is approximately half of the overall change of $\alpha$ with temperature. Note that the low-pressure, low-temperature $\alpha$ value of 2.59$\times 10^{-2}$/GPa at $p\,$=\,0.21\,GPa (pressure run $\textrm p_\textrm{2,PCC}$) and $T\,$=\,10\,K agrees well with the value of 2.52$\times 10^{-2}$/GPa, which was determined from the $^4$He-gas measurements at 10\,K up to 0.05\,GPa ($\alpha$ could not be determined up to higher pressures in the $^4$He-gas experiments due to the solidification of the medium). Unfortunately, the solidification of Helium and limitations of the maximum pressure of the gas-pressure setup do not allow us to clearly pin down the exact pressure dependence of $\alpha$ over wider ranges of pressures and temperatures. However, we note that whereas the $\alpha(T)$ data from the $^4$He-gas experiments (see Fig.\,\ref{fig1}\,(c)) seems to be almost independent of pressure for high temperatures, reasonable extrapolations of the $^4$He-gas pressure $\alpha$ data down to lower temperatures below the solidification of the $^4$He pressure medium might suggest that the pressure dependence of $\alpha$ becomes more pronounced upon cooling. Overall, our analysis from combining the $^4$He-gas data with the piston-cylinder cell data, presented here, provides some strong indications that $\alpha$ shows some non-negligible pressure dependence for $T\,\lesssim\,$60\,K. Although the exact reason behind this observation is unknown for now, we speculate that the stronger $p$-dependence of $\alpha$ is related to a possible change of the dominating electron scattering mechanism across $T\,\sim\,$50\,K, since a plot of the temperature-dependent d$R$/d$T$ (see Fig.\,\ref{fig3}\,(a) inset) shows a broad maximum at $\sim\,$50\,K. 
\begin{figure}
	\includegraphics[width=0.7\textwidth]{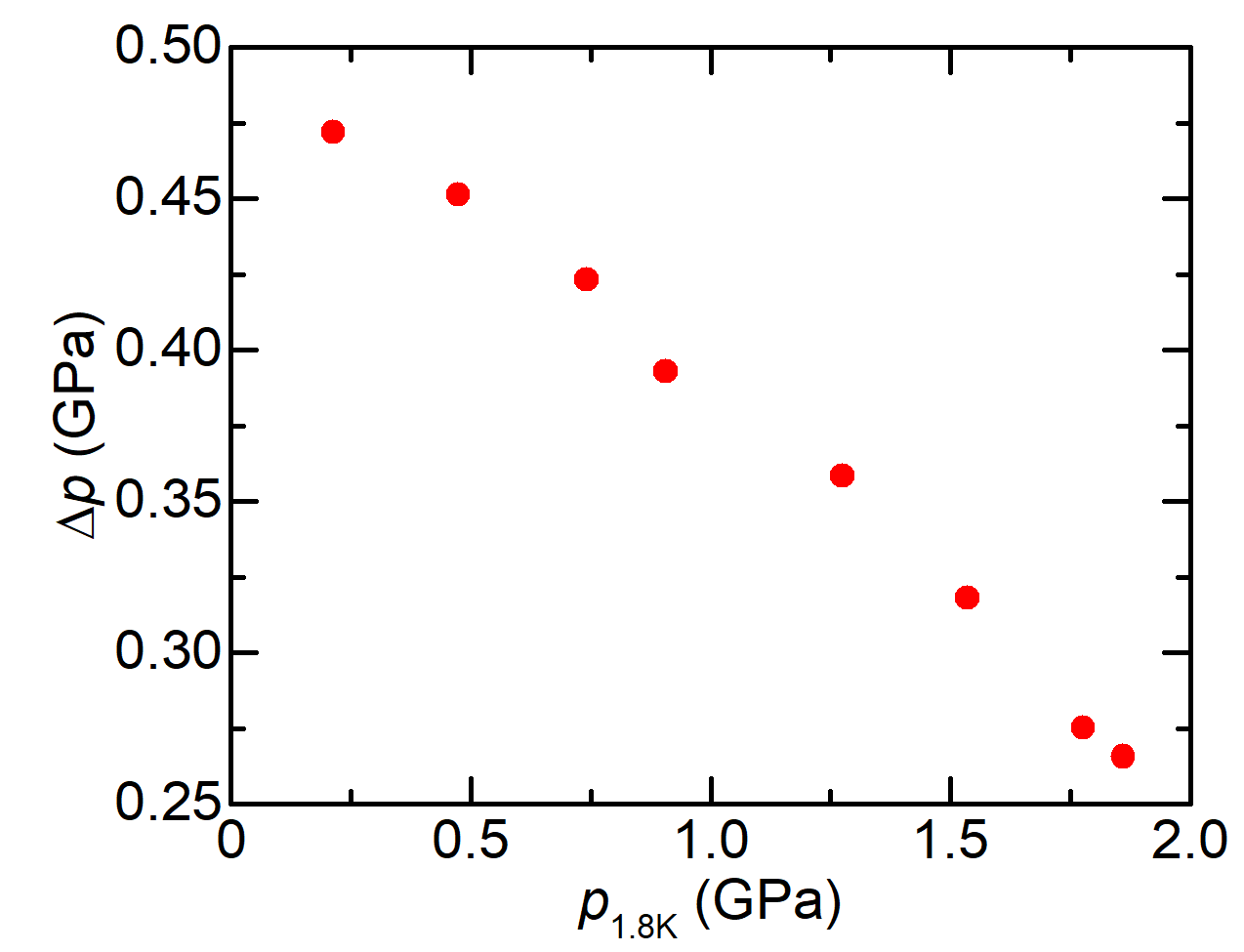}%
	\caption{Pressure drop between 300\,K and 1.8\,K, $\Delta p = p_\textrm{300\,K} - p_\textrm{1.8\,K}$, as a function of pressure in the piston-cylinder cell determined from the manganin manometer. Note that the modified $p(T)$ curves (dashed lines in Fig.\,\ref{fig4}) were used to determine $p_\textrm{1.8\,K}$. 
		\label{fig6}}
\end{figure}

Using the modified $p(T)$ data from the manganin sensor, we can now evaluate the pressure dependence of the pressure drop upon cooling from 300\,K to 1.8\,K, $\Delta p\,=\,p_\textrm{300\,K}-p_\textrm{1.8\,K}$, determined from the manganin sensor. As shown in Fig.\,\ref{fig6}, $\Delta p$ decreases upon increasing pressure, with $\Delta p\,\simeq$\,0.47\,GPa for $p_\textrm{1.8\,K}\,\simeq$\,0.21\,GPa and $\Delta p\,\simeq\,$0.26\,GPa for $p_\textrm{1.8\,K}\,\simeq\,$1.86\,GPa. These results are very close to earlier literature results, which found a pressure difference of $\sim$\,0.3\,GPa-0.4\,GPa between room temperature and liquid-nitrogen temperature for their specific pressure cells, media and sample space filling factors \cite{Itskevich1964,Brandt1974,Fujiwara1980,Thompson1984,Becker1976,Torikachvili2015}. Also, a previous study of $\Delta p$ in the same pressure cell with the same pressure medium \cite{Torikachvili2015} is consistent with our results in terms of the absolute values of $\Delta p$ as well as its pressure evolution.

We now compare the pressure values from the modified $p(T)$ curves (dashed lines in Fig.\,\ref{fig4}\,(a)) with those determined from elemental Pb (i.e., from the Pb-$T_\textrm c$ sensor) for low $T\,\sim\,7\,$K (solid circles). The Pb-$T_\textrm c$ sensor is frequently used in literature to infer the low-temperature pressure \cite{Smith1967,Smith1969,Clark1978,Wittig1979,Bireckoven1988}. Studies\cite{Smith1967,Smith1969,Clark1978,Eiling1981,Bireckoven1988} have shown that, upon increasing pressure up to $\sim$\,5\,GPa, the ambient pressure $T_\textrm c$\,=\,7.2\,K of Pb is suppressed linearly with a rate between $-$0.361\,K/GPa and $-$0.386\,K/GPa. By taking the suppression rate of -0.365\,K/GPa, as determined in Ref.\,\onlinecite{Eiling1981}, we determine the pressure at $T\,\sim$\,7 K and depict these pressures by solid circles in Fig.\,\ref{fig4}\,(a). The error bars for these data points are obtained using different pressure derivatives of Pb, reported in literature\cite{Smith1967,Smith1969,Clark1978,Eiling1981,Bireckoven1988}.

Overall, most of the pressure values from Pb-$T_\textrm c$ agree very well with those from the manganin sensor, using the extrapolation scheme outlined above (see Fig.\,\ref{fig4}\,(a)). This observation supports our modifications of the $p(T)$ curves obtained from manganin. On a more quantitative level, $p$ values at $T\,\sim$\,7\,K determined from manganin and Pb-$T_\textrm c$ differ by less than 0.025\,GPa for $p\,\lesssim\,$1.25\,GPa ($\textrm p_\textrm{2,PCC}$ to $\textrm p_\textrm{6,PCC}$). For $p\,\gtrsim$\,1.5\,GPa, the difference between pressure values inferred from the manganin and Pb-$T_\textrm c$ sensors becomes slightly larger, reaching $\sim$\,0.085\,GPa at 7\,K for our highest pressure run ($\textrm p_\textrm{9,PCC}$). The slightly larger difference of the pressure values for higher pressures could be due to the fact that the manganin sensor was only calibrated up to 0.8\,GPa in the $^4$He-gas pressure cell (maximum pressure of the system). Thus, any pressure dependence of $\alpha$ over a wider pressure range, even for $T\,>\,$60\,K, would directly affect the evaluation of the pressure from the magnanin sensor and therefore also its extrapolations.

Given that some small differences between pressure values determined from the manganin sensor and the Pb-$T_\textrm c$ are observed, we finally want to offer a practical approach for estimating the absolute pressure value at any given intermediate temperature for this specific combination of pressure cell, pressure medium and sample space filling factor. Since we lack any calibration measurements for the manganin sensor for higher pressures $p\,>\,$0.8\,GPa (due to the maximum pressure of our $^4$He gas setup), we suggest that if a Pb-$T_\textrm c$ manometer is present, one refers to the $p_\textrm{7K}$ obtained from $T_\textrm c$ of Pb for the determination of low-temperature pressure up to 2\,GPa. To estimate $p$ at higher temperatures, the $p(T)$ curves determined from manganin in this study can be used as a reference by using a linear interpolation of the nearest $p(T)$ curves, so that the interpolation matches $p_\textrm{7K}$ from Pb. If only a manganin sensor is present, $p(T)$ can be obtained by utilizing the $\alpha(T)$ characterized in $^4$He-gas experiments in this study (see Fig.\,\ref{fig2}\,(b)), and $p(T)$ at low temperatures ($T\,\lesssim\,60\,$K) can possibly be modified similar to the procedure performed in our analysis. This practical approach offered here gives a good estimation of the overall $p(T)$ behavior within the discussed systematic errors, which result from the small differences in the absolute values inferred from the manganin vs. the Pb-$T_\textrm c$ manometers. In general, we believe that a similar practical approach could be used to estimate pressure values at intermediate temperatures for other cells, pressure media and/or sample space filling factors as well by performing a separate calibration via a manganin sensor (and utilizing the $\alpha(T,p)$ behavior reported here) and a Pb-$T_\textrm c$ sensor.

\section{Summary}
In summary, so as to better characterize the temperature dependence of pressure within a piston-cylinder cell, the resistance of manganin for its use as a manometer was characterized in a $^4$He-gas pressure system from ambient pressure up to 0.8\,GPa and from room temperature down to the solidification temperature of $^4$He. Subsequently, the same manganin piece was measured in a piston-cylinder cell from ambient pressure up to $\sim\,$2\,GPa and from room temperature down to 1.8\,K. From an analysis of the resistance measurements, the temperature and pressure dependence of the pressure coefficient $\alpha(T,p)$ was determined. The abtained $\alpha(T,p)$ of manganin was utilized to study the temperature-dependent pressure behavior in a piston-cylinder cell and was compared to the low-temperature pressure, inferred from the superconducting transition temperature of elemental Pb. Our results can be summarized as follows: First, we find that $\alpha$ of manganin is 2.21$\times 10^{-2}$/GPa at 300\,K, which is in the range of other literature reports, and that $\alpha$ has a non-monotonic temperature dependence. Upon cooling, $\alpha(T)$ first decreases and then increases, thus displaying a broad minimum at $\sim\,$120\,K. In addition, our results suggest that $\alpha$ is almost pressure-independent for 60\,K\,$\lesssim\,T\,<\,$300\,K, whereas for $T\,\lesssim\,$60\,K it has a non-negligible pressure dependence, i.e., $\alpha$ at a given temperature is suppressed upon increasing pressure. Second, we quantified the $p(T)$ behavior in a piston-cylinder cell. We demonstrate that pressure decreases continuously upon cooling for the whole pressure range up to 2\,GPa, and that pressure experiences a more significant drop upon cooling through the medium solidification temperature. The difference in pressure between room temperature and low temperatures decreases upom increasing overall pressure. The low-temperature pressure values inferred from manganin are overall consistent with the ones inferred from the superconducting transition temperature of elemental Pb.

Overall, this work therefore provides two findings, which are important for the pressure community in general. First, we demonstrate that the temperature and pressure dependence of $\alpha(T)$ for manganin has to be taken into account for an accurate determination of $p(T)$ when using manganin as a manometer. Second, we provide a detailed analysis of the $p(T)$ behavior in piston-type pressure cells, which can be readily used in future pressure experiments to estimate the pressure at any given temperature. Whereas this work is done for a specific combination of pressure cell, pressure medium and sample space filling factor, we believe that our results can be used as reference to estimate pressure values at intermediate temperatures for piston-cylinder pressure cells with similar designs. For a more accurate and detailed $p(T)$ behavior analysis in other cells, for other used pressure media and/or other samples space filling factors, a separate calibration is needed, for which our generic analysis of $\alpha(T,p)$ of the manganin sensor will be useful.


\begin{acknowledgements}
We thank O. Palasyuk for making ribbons of material with chemical composition of Zeranin that were tested as potential resistive manometer. Work at Ames Laboratory was supported by the US Department of Energy, Office of Science, Basic Energy Sciences, Materials Sciences and Engineering Division. Ames Laboratory is operated for the US Department of Energy by Iowa State University under Contract No. DEAC0207CH11358. L.X., E.G. and R.A.R. were supported in part by the Gordon and Betty Moore Foundation’s EPiQS Initiative through Grant No. GBMF4411. Work at Frankfurt was supported by the DFG via the SFB/TR 288.
\end{acknowledgements}

\section{Data Availability}
The data that support the findings of this study are available from the corresponding author upon reasonable request.

\section{Appendix}

\subsection{Analysis of pressure run $\textrm p_\textrm{1,PCC}$ in the piston-cylinder cell}
\label{subsec:A}

\begin{figure}
	\includegraphics[width=0.7\textwidth]{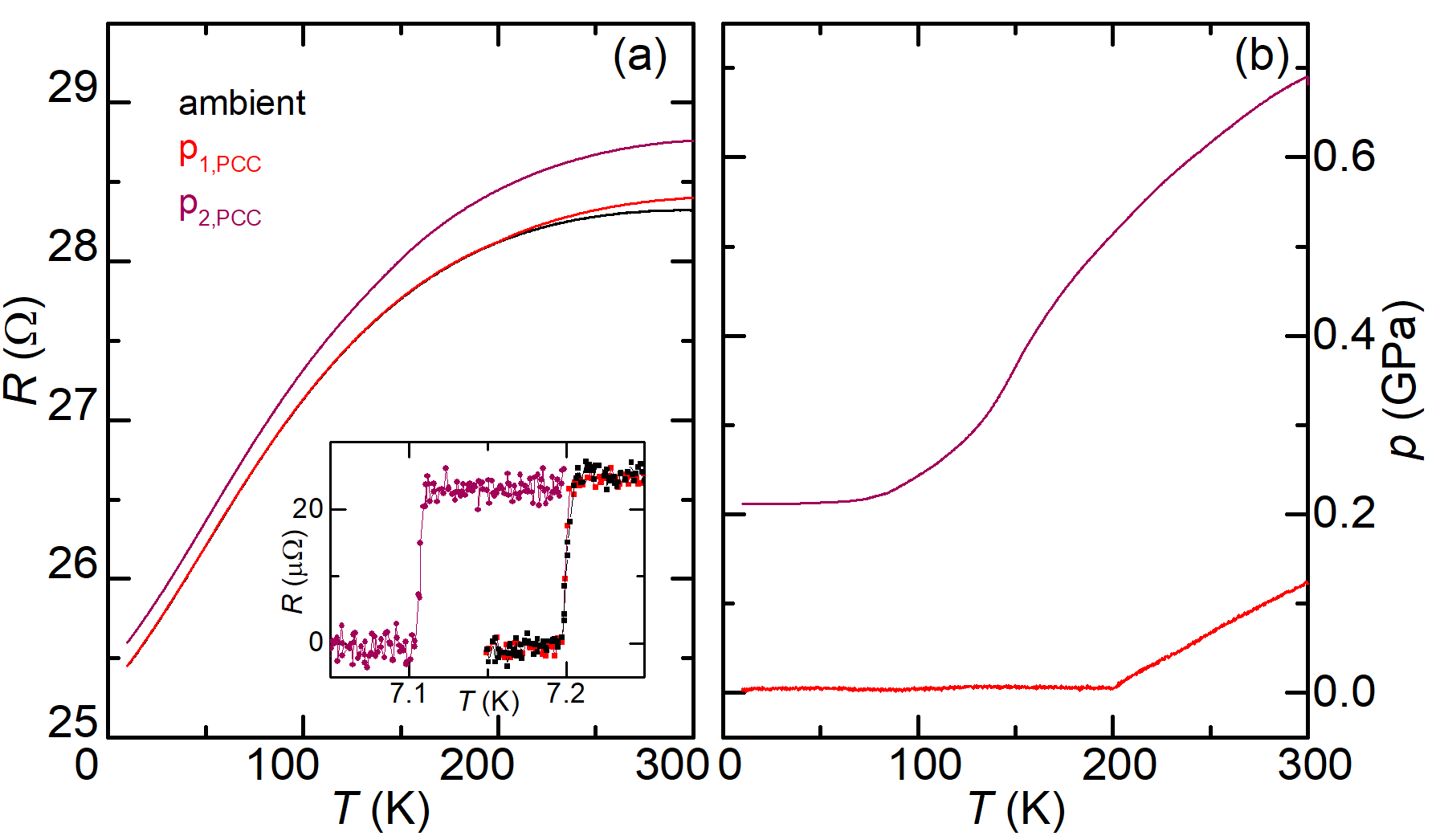}%
	\caption{(a) $R(T)$ of manganin measured at ambient pressure, $\textrm p_\textrm{1,PCC}$ (inside the piston-cylinder cell, for which the lock-nut was closed hand-tight without the application of external force) and $\textrm p_\textrm{2,PCC}$ (inside the piston-cylinder cell and first measurement, for which a finite force of $\sim$ 1000 lbs was applied to the piston prior to the measurement). Inset: $R(T)$ of Pb showing the superconducting transition for ambient pressure, for $\textrm p_\textrm{1,PCC}$ and for $\textrm p_\textrm{2,PCC}$; (b) Temperature-dependent pressure, $p(T)$, for $\textrm p_\textrm{1,PCC}$ and $\textrm p_\textrm{2,PCC}$ determined from $R(T)$ and $\alpha(T)$ of manganin.
		\label{S1}}
\end{figure}

Figure\,\ref{S1}\,(a) presents the temperature-dependent resistance of manganin at ambient pressure outside of the pressure cell, as well as inside the pressure cell without applying a load to the piston (``hand-tight'', $\textrm p_\textrm{1,PCC}$) and inside the pressure cell with a load of 1000 lbs applied to the piston that is locked by tightening the lock-nut ($\textrm p_\textrm{2,PCC}$). From the ambient pressure run to the $\textrm p_\textrm{1,PCC}$ run, resistance at any temperature above $\sim$\,200\,K increases. In contrast, no change of the resistance can be observed between the ambient and the $\textrm p_\textrm{1,PCC}$ run for temperatures below 200\,K. From $\textrm p_\textrm{1,PCC}$ to $\textrm p_\textrm{2,PCC}$, resistance increases at any temperature with increasing pressure. Figure\,\ref{S1}\,(a) inset shows the resistance of Pb across $T_\textrm c$ for the ambient pressure, $\textrm p_\textrm{1,PCC}$ and $\textrm p_\textrm{2,PCC}$ runs. We find that $T_\textrm c$ is the same for the ambient-pressure and the $\textrm p_\textrm{1,PCC}$ run, whereas it is distinctly lower for $\textrm p_\textrm{2,PCC}$. These data suggest that for $\textrm p_\textrm{1,PCC}$, the pressure at high temperatures is non-zero but becomes zero at low temperatures. We further calculated the temperature-dependent pressure for the $\textrm p_\textrm{2,PCC}$ run from manganin following the procedure outlined in the main text. As shown in Fig.\,\ref{S1}\,(b), a pressure of 0.12\,GPa is obtained at 300\,K for $\textrm p_\textrm{1,PCC}$. Upon cooling pressure decreases, reaches zero at $\sim$\,200\,K and apparently stays unchanged upon further cooling. A very similar result can be reached by using our ``practical'' approach to determining pressure as well. If we simply shift the $\textrm p_\textrm{2,PCC}$ curve down to 0.12\,GPa at 300\,K, we find that it crosses $p$\,=\,0 at $\sim$\,200\,K. In the main text, we demonstrated that a pressure loss of 0.47\,GPa occurs for $\textrm p_\textrm{2,PCC}$ upon cooling. Thus, when the room-temperature pressure is less than 0.47\,GPa, such as for $\textrm p_\textrm{1,PCC}$, the pressure will drop to zero already at an intermediate temperature (200\,K for $\textrm p_\textrm{1,PCC}$). We note that this might result in a inhomogeneous pressure for lower temperatures, since the differential thermal expansion between, e.g., 200\,K and low temperatures is still significant. Correspondingly, a minimum pressure of about 0.47\,GPa at room temperature is needed to guarantee a well-defined pressure environment down to lowest temperatures.

\subsection{Determination of pressure values in piston-cylinder cell via Pb-resistance manometer}

\begin{figure}
	\includegraphics[width=0.7\textwidth]{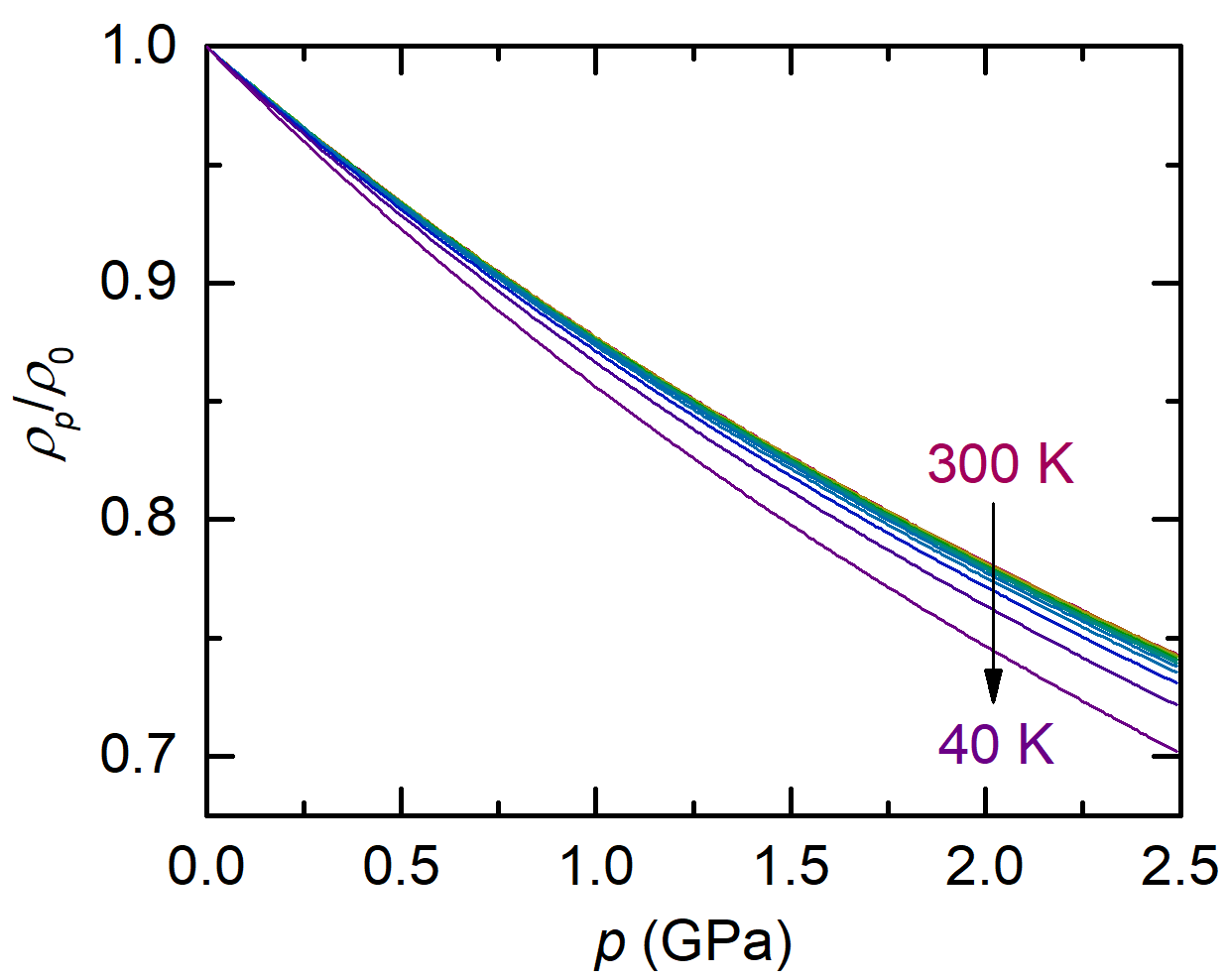}%
	\caption{Theoretical pressure-dependent resistivity of Pb normalized by ambient-pressure resistivity, $\rho_p/\rho_0$, at various fixed temperatures from 300\,K to 40\,K (spacing of 20\,K), based on the Bloch-Gr\"uneisen model of Ref.\,\onlinecite{Eiling1981}.
		\label{S5}}
\end{figure}

Similar to the manganin manometer, the resistance of Pb can be utilized to calculate pressure values as well (referred to here as Pb-resistive manometer). A. Eiling and J.\,S. Schilling in Ref.\,\onlinecite{Eiling1981} investigated the temperature and pressure dependence of resistivity of Pb and utilized the resistivity of Pb to calculate the pressure values in pressure cells\cite{Eiling1981}. We followed the analysis suggested in Ref.\,\onlinecite{Eiling1981} to carry out a similar determination of temperature-dependent pressure, $p(T)$, in the piston-cylinder cell from the Pb resistance data, which was measured in the present study (see Fig.\,\ref{fig3}\,(b) inset) simultaneous to the manganin resistance. The determined pressure values are compared with those from the manganin manometer and the Pb-$T_\textrm c$ manometer.

\begin{figure}
	\includegraphics[width=0.7\textwidth]{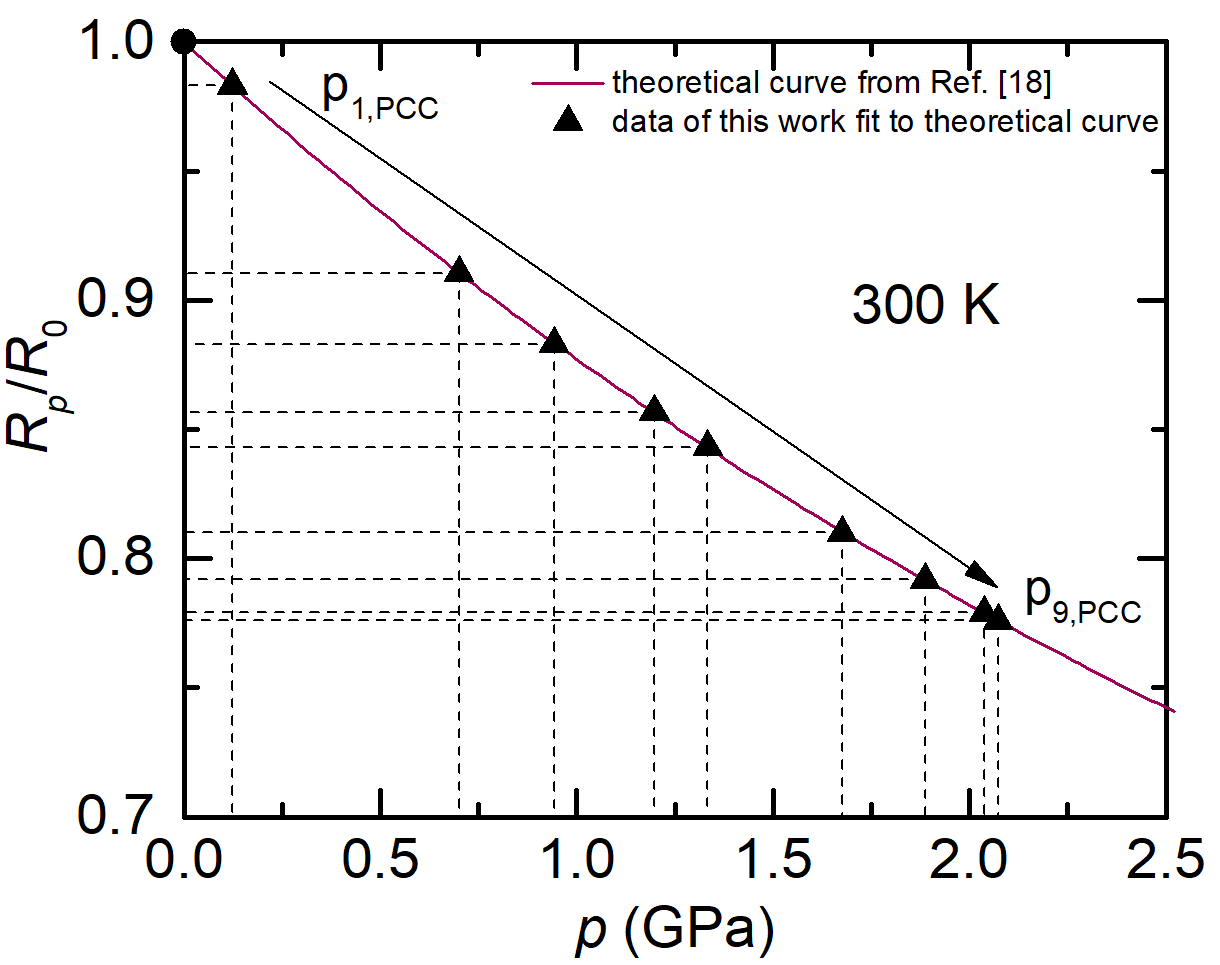}%
	\caption{Theoretical data of relative resistance, $R_p/R_0$, of Pb versus pressure at 300\,K (solid line), which is reprinted from Ref.\,\onlinecite{Eiling1981}, and experimental data from this work (triangles).  The circle symbol represents the data at ambient pressure which by definition is at $p$ = 0 GPa and $R_p/R_0$ = 1 in the plot. The experimental relative resistance data, is used to calculate pressure values at 300\,K for different pressure runs via fitting to the theoretical line.
		\label{S2}}
\end{figure}

\begin{figure}
	\includegraphics[width=0.7\textwidth]{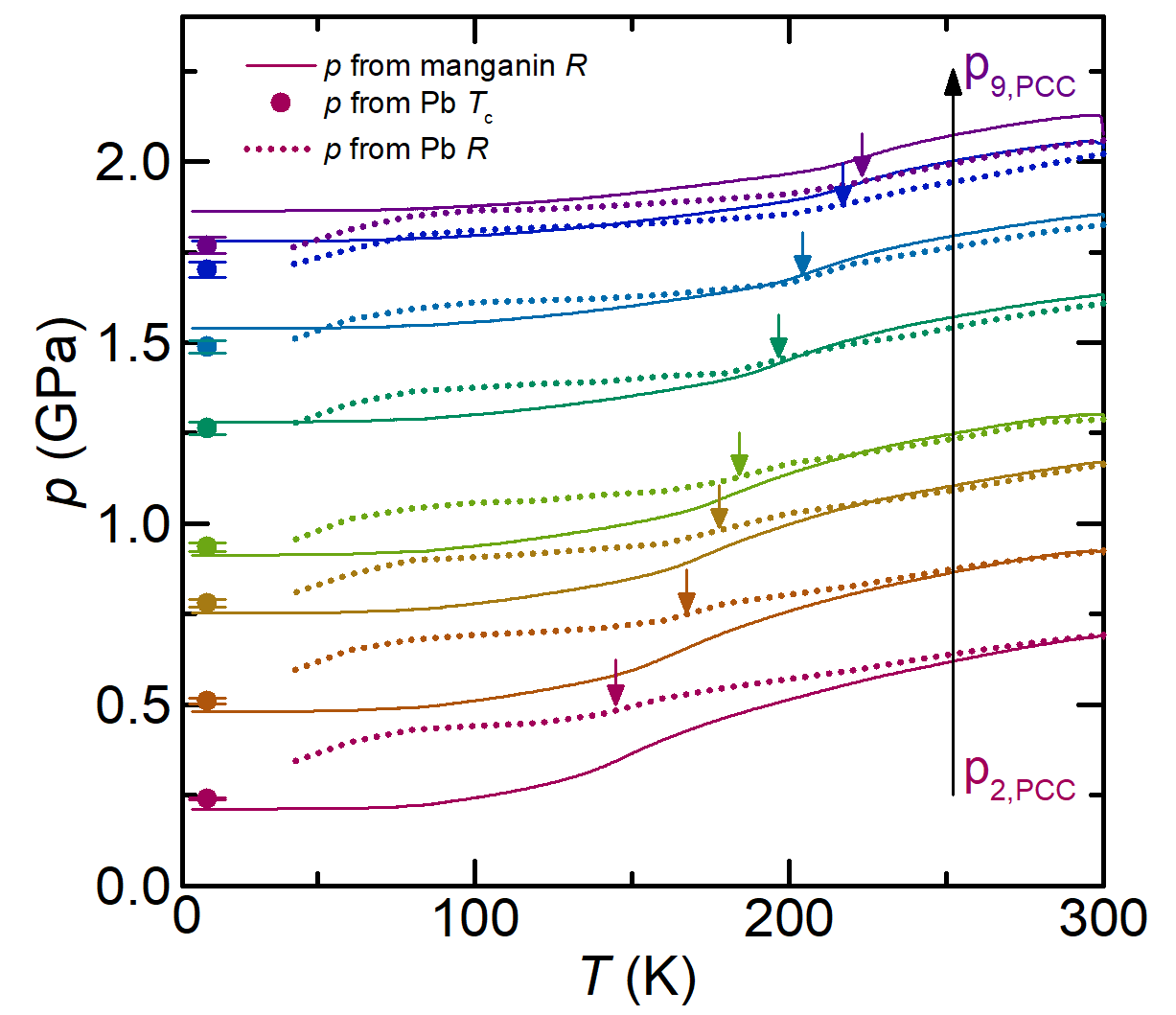}%
	\caption{Temperature-dependent pressure, $p(T)$, for various pressure runs in piston-cylinder cell. Dotted lines are $p(T)$ curves determined from Pb-resistive manometer, circles correspond to pressure values determined from the Pb-$T_\textrm c$ manometer and solid lines are (modified) $p(T)$ curves determined from manganin manometer. Downward arrows indicate a more rapid pressure decrease in $p(T)$ curves which is associated with the solidification of the pressure medium (see text for details).
		\label{S6}}
\end{figure}

According to the Bloch-Gr$\ddot {\textrm u}$neisen analysis outlined in Ref.\,\onlinecite{Eiling1981}, resistivity of Pb as a function of temperature and pressure, $\rho_p (T)$, can be calculated in the temperature range 7\,K$\,\le\,T\,\le\,$300\,K and pressure range of 0\,GPa$\,\le\,p\,\le\,$10\,GPa. Example theoretical curves of the resistivity, normalized by ambient-pressure resistivity, $\rho_p/\rho_0$, as a function of pressure at constant temperature are shown in Fig.\,\ref{S5}. These $\rho_p/\rho_0$ curves can be used to fit the measured experimental data, $R_p/R_0$ (assuming that the geometric dimensions of the Pb manometer do not change, in which case $R_p/R_0$ =  $\rho_p/\rho_0$), to calculate the pressure values. Figure\,\ref{S2} illustrates this procedure using the room-temperature data as an example. The solid line in Fig.\,\ref{S2} represents the theoretical $\rho_p/\rho_0$ ($R_p/R_0$) curve based on Ref.\,\onlinecite{Eiling1981} for $T\,=\,300$\,K, whereas the solid symbols represent the 300\,K experimental data obtained in this study for different pressure runs. Pressure values at 300\,K are calculated by fitting the experimental $R_p/R_0$ values to the theoretical curve. The same procedure was carried out for temperatures between 300\,K and 40\,K. Below $\,\sim\,$40\,K (about half of the Debye temperature), it was suggested in Ref.\,\onlinecite{Eiling1981} that the Bloch-Gr$\ddot {\textrm u}$neisen model becomes unreliable. The resulting $p(T)$ curves for various pressure runs ($\textrm p_\textrm{2,PCC}$ to $\textrm p_\textrm{9,PCC}$) are plotted in Fig.\,\ref{S6} together with those determined from the manganin and Pb-$T_\textrm c$ manometers. The $p(T)$ curves from the Pb-resistive manometer (dotted lines in Fig.\,\ref{S6}) manifest a continuous decrease of pressure upon cooling. A clear feature in $p(T)$ (as shown by the downward arrows in Fig.\,\ref{S6}) is associated with the solidification of the pressure medium\cite{Torikachvili2015}. Below $\sim$\,80\,K, a rapid decrease of $p$ upon cooling is observed. Such a rapid decrease of $p$ at low temperatures appears unphysical, following the same arguments, provided in the discussion of the non-modified $p(T)$ curves of manganin in the main text. We assume that this decrease can partially be attributed to a breakdown in the Bloch-Gr$\ddot {\textrm u}$neisen modeling of the Pb resistivity at low temperatures. Compared to the pressure values determined from other manometers, $p(T)$ curves from the Pb-resistive manometer (dotted lines) show a slower decrease of $p$ upon cooling (i.e., a smaller d$p$/d$T$) compared with those determined from manganin (solid lines) for $T\,\gtrsim$\,80\,K. In addition, extrapolations of Pb-resistive $p(T)$ either from above 80\,K or from below 80\,K down to 7\,K result in some discrepancies to the $p$ values determined from the Pb-$T_\textrm c$ sensor.

\begin{figure}
	\includegraphics[width=0.7\textwidth]{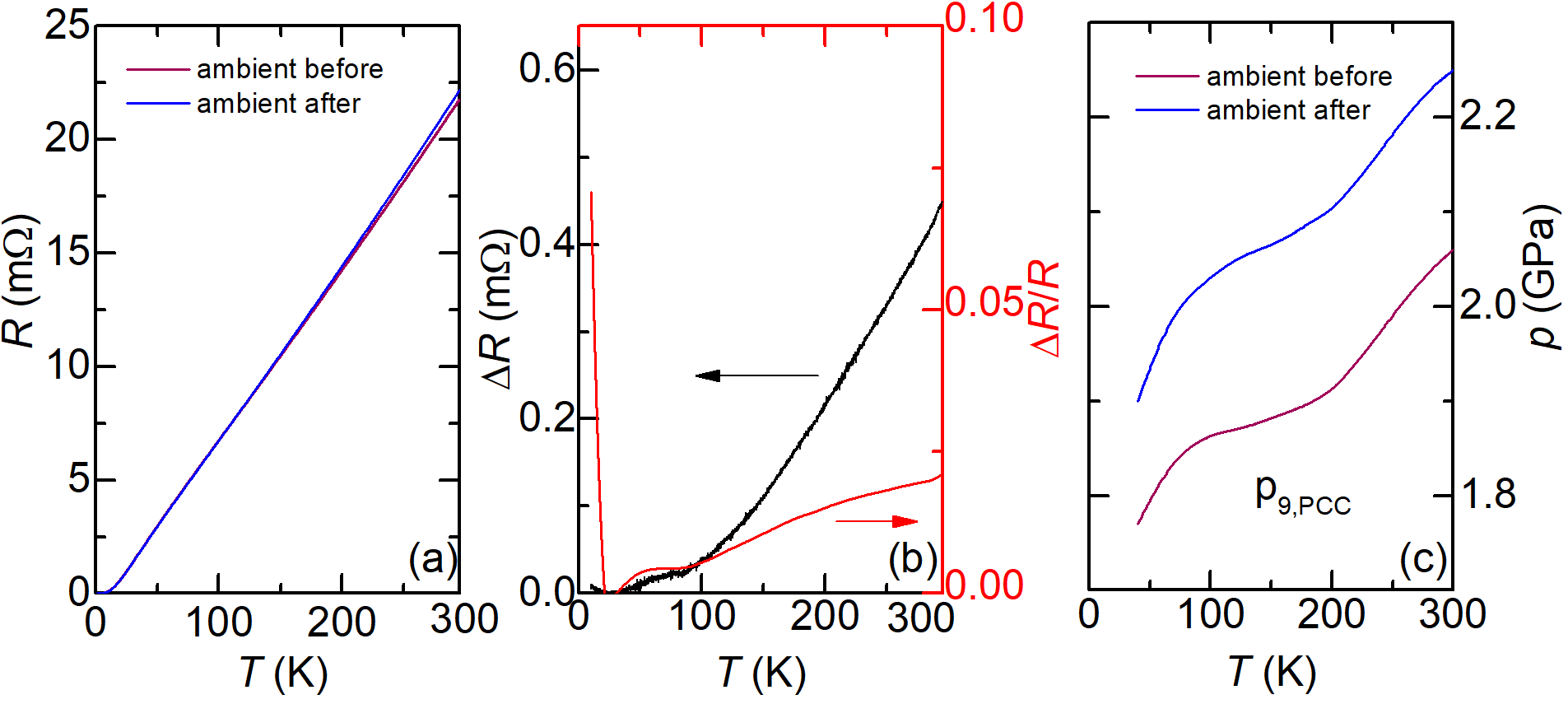}%
	\caption{(a) Temperature-dependent resistance of Pb at ambient pressure before and after pressure runs $\textrm p_\textrm{1,PCC}$ to $\textrm p_\textrm{9,PCC}$. Data curves are labeled as ``ambient before'' and ``ambient after'', respectively; (b) The resistance change, $\Delta R$ (left axis), between ``ambient before'' and ``ambient after'', as well as the relative change, $\Delta R/R$. where for $R$ the ``ambient before'' data was used (right axis), as a function of temperature; (c) Temperature dependence of pressure, $p(T)$, for pressure run $\textrm p_\textrm{9,PCC}$ using ``ambient before'' and ``ambient after'' data, respectively. See text for details.
		\label{S8}}
\end{figure}

In fact, inferring $p(T)$ from the Pb-resistive manometer is somewhat fraught with problems related to the residual resistivity of a sample as well as the the reproducibility of ambient-pressure resistivity values. To be more explicit, the $p(T)$ inferred from the Bloch-Gr$\ddot {\textrm u}$neisen analysis outlined above can vary depending upon the residual resistivity (RRR). Since $p(T)$ is inferred from $R_p/R_0$ (see Fig.\,\ref{S2}), changes in RRR affect the inferred $p(T)$.  For example, our initial Pb sample has a residual resistivity ratio, RRR $\sim$ 80, at ambient pressure; if we add a relatively small additional residual resistance (0.2 m$\Omega$) to change the RRR to 8, we find that pressure decreases more rapidly below $T\,\lesssim$\,80\,K for RRR\,=\,8, whereas the pressure decreases more moderately for $T\,\gtrsim\,80\,$K. Another problem is associated with pressure-induced changes of the ambient-pressure resistance, resulting from, e.g., changes in geometry and perfection of the Pb-resistive manometer. In Fig.\,\ref{S8}\,(a) we show the ambient-pressure resistance of the same Pb piece before and after the pressure runs. As can be seen, there is a non-trivial change in the resistance (see also $\Delta R$ and $\Delta R/R$ in Fig.\,\ref{S8}\,(b), where $\Delta R$ is the difference between the two ambient-pressure Pb resistance data sets). This change cannot simply be related to changes of geometry and cannot simply be related to changes of the defect scattering contribution. Figure\,\ref{S8}\,(c) shows the inferred $p(T)$ for the highest pressure run $\textrm p_\textrm{9,PCC}$ for 40\,K\,$\le\,T\,\le$\,300\,K, using the two sets of ambient-pressure resistance before and after the pressure cycle. There is a clear, $\sim$\,0.2\,GPa pressure difference at room temperature between the two $p(T)$ curves that becomes slightly smaller at lower temperatures. We point out that neither the residual resistance nor the potential change of geometry of the Pb manometer is taken care of in the Bloch-Gr$\ddot {\textrm u}$neisen analysis, and that any analysis in terms of the Bloch-Gr$\ddot {\textrm u}$neisen model is complicated by potential pressure-induced changes of the ambient-pressure resistivity of Pb. In summary, then, the use of Pb-resistive manometer for determining $p$ values over a wide temperature range is associated with larger uncertainties than the use of the manganin sensor and as such was not used in the main text. This said, it is very important to note that the Pb-$T_\textrm c$ value is not affected by these concerns (i.e., changes of RRR, changes of geometry or general changes of ambient-$p$ resistance) and is therefore a much more robust manometer when measuring the pressure values at low temperatures.

\clearpage

\bibliographystyle{apsrev4-1}
%
\end{document}